# Elucidation of molecular targets of bioactive principles of black cumin relevant to its anti-tumour functionality - An *Insilico* target fishing approach

Amulyashree Sridhar[1], Sadegh Saremy[2*], Biplab Bhattacharjee[1]

[1]Department of Bioinformatics, PES Institute of Technology, Bangalore, India

[2]Department of Biotechnology, Brindavan College, Bangalore, India

**Abstract:**

Black cumin (*Nigella sativa*) is a spice having medicinal properties with pungent and bitter odour. It is used since thousands of years to treat various ailments, including cancer mainly in South Asia and Middle Eastern regions. Substantial evidence in multiple research studies emphasizes about the therapeutic importance of bioactive principles of *N. sativa* in cancer bioassays; however, the exact mechanism of their anti-tumour action is still to be fully comprehended. The current study makes an attempt in this direction by exploiting the advancements in the Insilico reverse screening technology. In this study, three different Insilico Reverse Screening approaches have been employed for identifying the putative molecular targets of the bioactive principles in Black cumin (thymoquinone, alpha-hederin, dithymoquinone and thymohydroquinone) relevant to its anti-tumour functionality. The identified set of putative targets is further compared with the existing set of experimentally validated targets, so as to estimate the performance of *insilico* platforms. Subsequently, molecular docking simulations studies were performed to elucidate the molecular interactions between the bioactive compounds & their respective identified targets. The molecular interactions of one such target identified i.e. VEGF2 along with thymoquinone depicted one H-bond formed at the catalytic site. The molecular targets identified in this study need further confirmatory tests on cancer bioassays, in order to justify the research findings from Insilico platforms. This study has brought to light the effectiveness of usage of Insilico Reverse Screening protocols to characterise the un-identified target-ome of poly pharmacological bioactive agents in spices.

**Introduction:**

Spices and medicinal herbs are known to have anti-cancer characteristics which can be used to target the tumor growth and further damage [1]. One such spice is *Nigella sativa*, also called as black cumin. The spice *Nigella sativa* belongs to *Ranunculaceae* family. It is an annual herb widely seen in India, Middle East, Europe, Greece and Egypt [2]. It is listed in the "Medicine of the Prophet" as a natural remedy to cure all pathological conditions. Its active ingredients are used from ancient times as Unani, Ayurveda, and Chinese medication for various diseases like asthma, rheumatoid arthritis, immune and inflammatory diseases [3]. The main active component of black cumin volatile oil is Thymoquinone (TQ), of about 54% and others include monoterpenes like p-cymene, dithymoquinone ($TQ_2$), thymohydroquinone (THQ) and alpha-pinene [4]. The seeds are thought to contain active ingredients having anti-tumor characteristics [5].

A number of theories have been postulated concerning the anti-tumour mechanism of action of black cumin and its bioactive ingredients against cancer. The first theory states that Thymoquinone, a principle component in Black cumin inhibits the DNA synthesis process leading to apoptosis and thereby hindering the cell proliferation. 5-fluorouracil inhibits thymidine synthesis and is seen to cure many forms of cancer. A number of studies have established that the pro-apoptotic mechanism of thymoquinone is mediated by both p53 dependent pathways and p53 independent pathways. It is also observed that there is a substantial inhibition of Bcl-2 proteins, NF-kappa B, DNA methyl transferases, Histone deacetylase by thymoquinone [4]. The second theory postulated that the anti-tumour effects of black cumin is because of its potent anti-inflammatory properties. This was established by a research study on thymoquinone on pancreatic cells, in which it was observed that this compound inhibits NF-κβ pathway leading to reduction in the growth of pancreatic cancer [6]. The third theory postulated that the anti-cancer effect of black cumin is because of its anti-angiogenic mechanism. In this context, it was observed in *invitro* and *invivo* study that thymoquinone inhibits the AKT and EKT signalling pathways and depicts inhibitory action on key cancer regulatory proteins such as, VEGF, NF-κβ, c-Jun [7].

The process of identification of molecular targets for the established small molecules utilizes a series of biochemical and quantitative proteomics protocols. These techniques are time consuming and protocols needs to be standardized. In recent times, an alternative approach named "Insilico Reverse Screening "has become very popular for the task of molecular target identification. This method utilizes a computational search strategy on a panel of chemogenomics- derived target database to derive a set of putative targets for a given small molecule [8]. This approach has been successfully employed to decipher the molecular targetability of active ingredients used in Ayurveda medicine [9]. Reverse pharmacophore mapping is based on alignment of the ligand to an available pharmacophore from a panel of pharmacophore database. Many such tools have been developed in the recent years. They include Tarfisdock [10], PharmMapper [11], ReverseScreen3D [12], idTarget [13], Target Hunter [14] and ChemMapper [15].

In the current study, a unified approach integrating PharmMapper, ReverseScreen3D and TargetHunter to identify the potential targets for *N.sativa* bioactive ingredients. The targets prioritized were further validated for its anti-cancerous activity by matching up to experimental proven targets. Optimization is performed by studying the binding poses of these probable targets against the bioactive agents.

**Methodology:**

**Identification of bioactive constituents of Black cumin by Literature review**

Black cumin has number of chemically diverse bioactive agents contributing to its anti-cancer functionality. Literature review revealed the principal bioactive components of black cumin i.e. Thymoquinone, alpha-hederin, dithymoquinone and thymohydroquinone. The structures of these molecules were retrieved from NCBI PubChem in sdf format.

**Reverse Screening approaches to identify potential therapeutic targets**

The reverse screening strategy in this study has been implemented using three insilico platforms namely, PharmMapper, ReverseScreen3D and TargetHunter. The sdf and .mol2 formats of the bioactive constituents of N.sativa i.e. Thymoquinone, Alpha-hederin, Dithymoquinone and Thymohydroquinone were submitted to the three tools as an input. The PharmMapper gives the best mapping poses by comparing with available targets in PharmTargetDB and the respective N-best fit poses are generated. In ReverseScreen3d, for each target protein a single ligand that has highest similarity with 2D similarity of input molecule is generated, furthermore, a 3D similarity search is performed to match the ligands generated from the database with the input molecule. Target Hunter utilizes an algorithm for predicting the targets, namely the "Targets Associated with its MOst SImilar Counterparts to generate hits. The target list generated as output from the three tools was further annotated to screen out the possible target list having significant association to cancer.

**Retrieving experimentally validated target's information from Bioassay databases**

The targets identified from the reverse screening approach were further validated for their anti-cancer properties experimental repositories including NCBI PubChem, NPACT [16] and Cancer Resource database [17].

**Comparative analysis of identified putative targets with the experimental targets**

The potential targets identified from reverse screening approaches were compared with the experimental results to validate the results obtained from the inverse screening approaches.

**Classification of potential targets based on mode of action**

To understand the precise mechanistic behaviour of each bioactive agent, the identified targets were categorized into different functional classes: a) anti-proliferation b) anti-apoptosis c) anti-inflammation d) anti-invasive.

**Molecular Docking studies of the potential targets with the bioactive components**

Docking studies were performed to depict the protein and the bioactive agent interactions using AutoDock PyRx4.2. The 3D structures of identified targets were retrieved from RCSB PDB and the structures of small molecules were retrieved from NCBI PubChem. These molecules along with targets were submitted to AutoDock PyRx4.2 to perform docking simulations.

**Validation of the docked poses**

To re-affirm the successful target identification in the current study, molecular docking studies was conducted on Crystallographic structure having bound ligand. The structure of Vascular Endothelial Growth Factor-2 (VEGF-2) was retrieved from RSCB-PDB (PDB ID: 3CP9). The co-crystallized ligand was removed from the original structure and was re-docked.

**Results and Discussion:**

The targets identified for thymoquinone from insilico techniques are depicted in Table 1.

**Anti-proliferative and Anti-apoptotic mechanisms in *N.sativa***

The principle agents of N.sativa including Thymoquinone, Alpha-hederin and Dithymoquinone were found to have potent anti-proliferative and anti-apoptotic mechanisms. From the "set of putative targets" identified for Thymoquinone by Insilico Reverse Screening process, it was observed that many of these identified targets Cyclin-A2, Cell division protein kinase 2, Dihydrofolate reductase, Deoxycytidine kinase, Caspase 3, Bcl-X, Heat shock protein HSP-90 alpha, are significant contributor to the process of cell proliferation and apoptosis in cancer. Alpha-hederin also identified some prevalent targets including Ras related protein rap-2a, Cathepsin K, Estradiol 17-beta dehydrogenase-1, GTPase HRas, Cellular retinoic acid-binding protein 2, Dihydroorate dehydrogenase and many more.

**Anti-inflammatory mechanisms in *N.sativa***

The anti-inflammatory potency of Nigella sativa is predominantly contributed by its principle ingredient Thymoquinone. Using the Insilico Reverse Screening protocol it was identified that the putative targets of thymoquinone responsible for its anti-inflammatory action are the following: Aldose reductase, Leukotriene A-4 hydrolase, inducible nitrogen oxide synthase (iNOS), and Glutathione-requiring prostaglandin D synthase.

**Anti-angiogenesis mechanism in *N.sativa***

Using the Insilico Reverse Screening protocol it was identified that the putative targets of thymoquinone, alpha-hederin, dithymoquinone and thymohydroquinone responsible for its anti-angiogenic action are the following : VEGF, MMP-9, Macrophage metalloelastase, Stromelysin , Basic fibroblast growth factor.

**Molecular Docking analysis with VEGF 2**

VEGF is a cell signalling protein that initiates vasculogenesis and angiogenesis. It acts as supplier of oxygen during inadequate blood supply. Overexpression of VEGF expression leads to cancer followed by metastasis. The cellular response is mediated by binding to tyrosine kinase receptors called VEGFRs which dimerize and activate by trans phosphorylation. They are categorized as VEGFR-1, VEGFR-2 and VEGFR-3. It is seen that VEGF binding to VEGFR-2 domains 2 and 3 enhances the likelihood of other VEGFR-2 binding. It forms a dimer and they are stabilized followed by auto phosphorylation leading to activation of extracellular receptor kinase (ERK) signalling pathway.

VEGFR-2 tyrosine kinase domain has a dual 70 amino acid insert region for kinase. The dimer is formed when VEGF-A binds to VEGFR-2 followed by auto-phosphorylation at insert kinase region and carboxy terminal tail [18]. VEGFR-2 (PDB ID: 3CP9) protein used in this study for molecular docking simulation has an inbound ligand 3-(2-aminoquinazolin-6-yl)-1-(3,3-dimethylindolin-6-yl)-4-methylpyridin-2(1H)-one, referred as C19. It is an aminoquinazoline pyridine capable of inhibiting several targets like KDR, p38, Lck and Src [19]. The original crystallographic ligand was removed from the protein and re-docked to analyse the conformation difference of the ligand in pre-docking and post-docking poses. The pre-docked and post-docked poses were superimposed using PyMOL software.

The pre-docked pose of the ligand with the VEGF-2 retrieved from PDBsum indicated that the bound ligand formed 3 hydrogen bonds: one bond with Asp1046 (A) and two bonds with Cys919 (A). The post-docked pose (Seen in Figure 1(b)) indicated that original ligand (C19) formed 4 hydrogen bonds: one bond with Ile1044 (A), Asp1046 (A) and two bonds with Cys919 (A).

The H-bonds formed and RMSD (Root Mean Square Deviation) were calculated for the theoretical model as well as the experimental structure and it was observed that the H-bond length deviation between the crystal structure and docked pose was less than 0.20Å. The RMSD for C19 between the crystal structure and docked poses was 0.065Å. The results indicate ability of AutoDock simulations to replicate the experimental binding information of ligand-VEGF 2 complex.

**Binding pose of Thymoquinone with VEGF 2**

The post-docked binding conformation of thymoquinone with VEGF2 is illustrated in the Figure 1(b). It has a binding energy of -7.57 kcal/mol. The observations indicate that thymoquinone binds to VEGF-2 at the same catalytic site as that of the C19. Interaction of thymoquinone at the active site involves Leu840, Val848, Ala866, Val867, Glu885, Thr916, Leu1035, Cys1045 and Asp1046, which is analogous to the binding pattern of C19 at the catalytic site. Thymoquinone forms single N-H...O hydrogen bond with Asp1046 having bond length of 3.20Å. The similar hydrogen bond can be observed in the C19 with bond length of 3.08 Å. The interaction of thymoquinone is mainly hydrophobic. The docking cluster analysis of all runs showed a consistent interaction with Asp1046. The interactions are studied via the electrostatic pattern and hydrogen bonds formed. The results indicate capacity of thymoquinone to act as a potent inhibitor of VEGF2 at its catalytic site.

**Conclusion:**

Many spices have traditionally been used in culinary and therapeutic applications by traditional practitioners without being aware of their precise mechanism of action. The current study is a true attempt to elucidate the exact poly pharmacological targetability of bioactive agents of black cumin. The putative target set identified by the "Insilico target fishing" approach in this study correlate well with the experimental findings. The theories of the anti-cancer mechanism of bioactive agents of N. Sativa were further strengthened by the identification of targets from different categories including, anti-proliferative, anti-apoptosis, anti-inflammatory, and anti-angiogenic. Molecular docking studies of the "putative target set identified targets" with their respective bioactive compounds demonstrated their binding at the catalytic site of the protein targets, thereby proving their antagonistic mechanism. Further validation studies with VEGF-2 with thymoquinone demonstrated that the catalytic action is because of the inter-molecular H-bonds and other interactions. The "putative target set identified targets" has to be further validated in invivo and invitro bioassays to support the findings of this study.

**Acknowledgements**

The authors express their sincere gratitude to Department of Biotechnology, PESIT Bangalore for the continuous support during the work.

**Table 1**: The primary targets of thymoquinone identified from reverse screening approaches

| PDB ID | Protein Target | Procedure of reverse screening | Experimental Evidence | Binding energy (kcal/mol) |
|---|---|---|---|---|
| 2FGI | Basic fibroblast growth factor receptor 1 | PharmMapper/ ReverseScreen3D | CancerResource | -4.65 |
| 1P62 | Deoxycytidine kinase | PharmMapper | CancerResource | -4.75 |
| 2X0V | Cellular tumor antigen p53 | ReverseScreen3D | NPACT | -4.6 |
| 1DB1 | Vitamin D3 receptor | PharmMapper/ ReverseScreen3D | CancerResource | -5.17 |
| 1R9O | Cytochrome P450 2CP | PharmMapper/ ReverseScreen3D | CancerResource | -4.63 |
| 1E3G | Androgen receptor | PharmMapper/ ReverseScreen3D | CancerResource | -5.28 |
| 2C6T | Cyclin A2 | PharmMapper | CancerResource | -4.77 |
| 3BL1 | Carbonic Anhydrase II | PharmMapper/ ReverseScreen3D | CancerResource | -4.33 |
| 1HFQ | Dihydrofolate reductase | PharmMapper/ ReverseScreen3D | CancerResource | Error |
| 3CP9 | Vascular endothelial growth factor 2 | PharmMapper/ ReverseScreen3D | CancerResource | -7.57 |
| 1B41 | Acetylcholine esterase | Target Hunter | NCBI PubChem | -4.09 |
| 2QG0 | Heat shock protein HSP-90 alpha | PharmMapper | CancerResource | -4.14 |
| 3F7H | Baculoviral IAP repeat-containing protein 7 | PharmMapper/ ReverseScreen3D | CancerResource | -3.73 |
| 2FZ8 | Aldose reductase | PharmMapper/ ReverseScreen3D | CancerResource | -5.99 |
| 1RW8 | TGF-beta receptor type-1 | PharmMapper/ ReverseScreen3D | CancerResource | -5.21 |
| 1GRE | Glutathione reductase, mitochondrial | PharmMapper | CancerResource | -4.65 |
| 2R7B | 3-phosphoinositide-dependent protein kinase 1 | PharmMapper/ ReverseScreen3D | CancerResource | -4.59 |
| 1NME | Caspase 3 | ReverseScreen3D | CancerResource | -3.99 |
| 1YSG | Apoptosis regulator Bcl-X | ReverseScreen3D | NPACT | -3.89 |
| 3PXZ | Cell division protein kinase 2 | PharmMapper, ReverseScreen3D | CancerResource | -5.21 |
| 1GKC | Matrix metalloprotease 9 | PharmMapper, ReverseScreen3D | CancerResource | -5.64 |
| 2RG6 | MAP kinase 14 | PharmMapper, ReverseScreen3D | CancerResource | -4.81 |

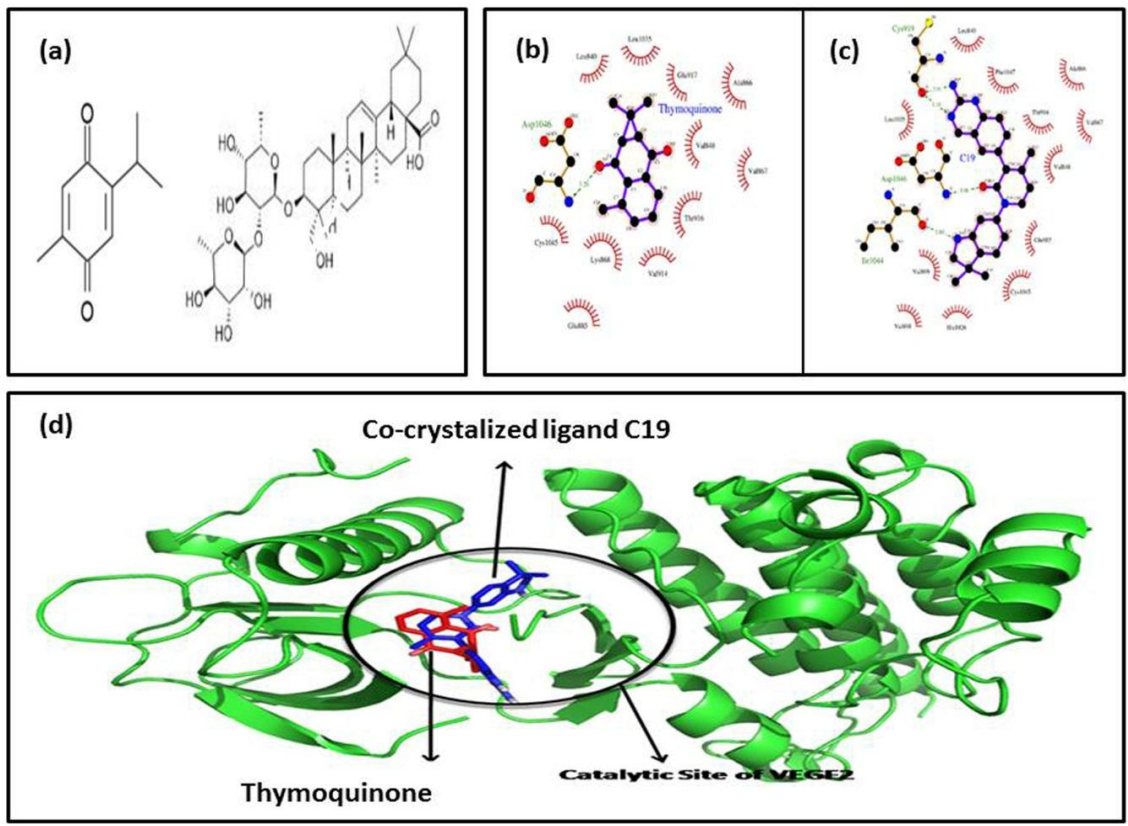

**Figure-1**: (a) The molecular structures of bioactive agents of *N.sativa* namely, Thymoquinone and alpha-hederin; (b) The binding site of Thymoquinone and C19 to VEGF2 depicted in microenvironment. Thymoquinone binds with a single H-bond with Asp1046, whereas in (c) the original ligand, C19 forms 4 hydrogen bonds with residues Asp1046, Ile1044 and two hydrogen bonds with Cys919 (green dotted lines); (d) The docked conformation of Thymoquinone and the original ligand, C19 with VEGF2 chain A.